\documentclass[english]{rspublic}
\usepackage[latin9]{inputenc}
\usepackage{textcomp}
\usepackage{graphicx}
\usepackage{amssymb}

\newcommand{\lyxaddress}[1]{
\par {\raggedright #1
\vspace{1.4em}
\noindent\par}
}

\usepackage{babel}

\begin{document}

\title{How to characterize the dynamics of cold atoms in non dissipative
optical lattices?}

\author{D. Hennequin and Ph. Verkerk}

\maketitle

\lyxaddress{Laboratoire PhLAM, UMR CNRS, CERLA, Université de Lille 1, 59655
Villeneuve d'Ascq, France}

\textbf{\footnotesize Abstract. }{\footnotesize We examine here the
classical dynamics of cold atoms in square optical lattices, i.e.
lattices obtained with two orthogonal stationary plane waves. Contrary
to much of the past studies in this domain, the potential is here
time independent and non dissipative. We show that, as a function
of the experimental parameters, very different behaviors are obtained,
both for the dynamics of atoms trapped inside individual sites, and
for atoms travelling between sites: inside the sites, chaos may be
a main regime or, on the contrary, may be negligible; outside the
sites, chaos sometimes coexists with other regimes. We discuss what
are the consequences of these differences on the macroscopic behavior
of the atoms in the lattice, and we propose experimental measurements
able to characterize these dynamics and to distinguish between the
different cases.}{\footnotesize \par}

\section{Introduction}

The cooling of atoms to extremely low temperatures, through the use
of magneto-optical traps (MOT), opened since the mid eighties fantastic
possibilities to increase our experimental knowledge of the quantum
world. The most spectacular realization was the achievement of the
Bose-Einstein condensation, and thus of macroscopic quantum objects.
However, even in the classical world, the possibility to study the
dynamics of atoms not {}``blurred'' by the Doppler effect is very
exciting. This requires to develop tools to manipulate the atoms,
for e.g. guiding them or {}``designing'' their phase space.

Optical lattices provide such tools: their versatility allows to manipulate
atoms with an extreme precision and a relative ease (Guidoni \& Verkerk
1999). Because of these qualities, they represent an outstanding toy
model, and have recently attracted increasing interest in various
domains. Condensed matter systems and strongly correlated cold atoms
in optical lattices offer deep similarities, as in the superfluid-Mott
insulator quantum phase transition (Greiner \textit{et al.} 2002),
or in the Tonks-Girardeau regime (Paredes \textit{et al.} 2004). Here,
the interactions between atoms play a crucial role, and require the
use of a Bose-Einstein condensate. In particular, instabilities are
expected in the Gross-Pitaevskii equation, because of the non-linear
term (Thommen \emph{et al.} 2003, Fang \& Hai 2005, Kuan \emph{et
al.} 2007). Quantum computing also requires a coupling between atoms:
the optical lattices appear to be an efficient implementation of a
Feynman's universal quantum simulator (Jaksch \& Zoller 2005), and
are among the most promising candidates for the realization of a quantum
computer (Mandel \textit{et al.} 2003, Vollbrecht \textit{et al.}
2004). On the other hand, noninteracting atoms also exhibit interesting
behaviors. In this case, the physics is essentially that of a single
atom. A higher number of atoms simply increases the observable signal.
That is the case in statistical physics, where cold atoms in optical
lattices, through their tunability, made possible the observation
of the transition between Gaussian and power-law tail distributions,
in particular the Tsallis distributions (Douglas \textit{et al.} 2006,
Jersblad \emph{et al.} 2004). or that of Anderson localization (Billy
\textit{et al.} 2008, Roati \emph{et al.} 2008, Chabe \textit{et al.}
2009).

Non interacting cold atoms appear also to be an ideal model system
to study the dynamics of a system in its classical and quantum limits.
Both are closely related, as the latter is only defined as a function
of the former. In particular, quantum chaos is defined as the quantum
regime of a system whose classical dynamics is chaotic. A good understanding
of the classical dynamics is therefore an essential prerequisite to
the study of quantum dynamics. In non dissipative optical lattices,
both the classical and the quantum limits are experimentally accessible,
and it is even possible to change quasi continuously from a regime
to the other (Steck\textit{ et al.} 2000). Moreover, the extreme flexibility
of the optical lattices makes it possible to imagine a practically
infinite number of configurations by varying the complexity of the
lattice and the degree of coupling between the atoms and the lattice.

Many results have been obtained during these last years in the field
of quantum chaos (Steck\textit{ et al.} 2000, Lignier\textit{ et al.}
2005). However, all these works have used very simple potentials,
mainly 1D. Chaos is obtained only with a periodic (or quasi-periodic)
temporal forcing of the depth of the lattice (Steck\textit{ et al.}
2000, Lignier\textit{ et al.} 2005), and only the temporal dynamics
of the individual atoms is studied. The introduction of this external
clock and the restriction to 1D potentials reduce considerably the
generality of these results and the type of possible dynamics. In
particular, the behaviors related to the appearance of new frequencies
or to a frequency shift (quasi-periodic and homoclinic bifurcations,
for example) are impossible.

If we want to break these limitations, several problems have to be
examined: what type of time-independent lattice will lead to a reasonably
complex dynamics? What are the relevant quantities to characterize
this dynamics? And what are those which can be implemented experimentally?
Note that these questions need first to be answered for the \emph{classical}
atoms. We have to search for a configuation leading to complex \emph{classical}
dynamics. And experimentally, the dynamics in the \emph{classical}
limit must be characterized before considering the quantum system.
In this paper, we address these questions, limiting our analysis to
the classical limit. The quantum counterpart will be discussed in
a future work. In section 2, we give some facts about cold atoms and
optical lattices for those who are not familiar with this domain,
and we discuss of possible lattices leading to complex dynanamics.
Section 3 is devoted to the dynamics of atoms inside the wells, while
section 4 deals with the dynamics of atoms traveling between several
wells. Finally, we discuss in section 5 of the possible implementation
of experimental measurements.

\section{Context: cold atoms and optical lattices.}

Cold atoms refer here to atoms cooled through a magneto-optical trap
(MOT). The cooling is mainly obtained through an exchange of the momentum
between an atom and a counter-propagating optical beam: while the
absorption of a photon by the atom leads to a deceleration of the
atom in the direction of the beam, the spontaneous re-emission of
the photon arises in a random direction, and so does not change, in
average, the atom velocity. To slow down atoms in 3D, three pairs
of counterpropagating laser beams are necessary. Obviously, a moving
atom is decelerated by the photons traveling in a direction opposite
to its own, but is accelerated by the photons traveling in the same
direction as its own. But the frequency of these trap beams is detuned
to the red of the atomic transition, so that, because of Doppler effect,
the front photons are closer to resonance, and thus the deceleration
process is more efficient than the acceleration one. This Doppler
cooling process is coupled to an inhomogeneous magnetic field, which
enhances the cooling process through the Zeeman levels splitting,
and adds a restoring force to increase the atomic density of the cloud
of cold atoms. MOTs lead typically, for Cesium atoms, to a 2 mm diameter
cloud of $10^{8}$ atoms at $5\:\mu\mathrm{K}$. Such a cloud of cold
atoms can exhibit spatio-temporal instabilities and chaos (Wilkowski
\textit{et al.} 2000, di Stefano \textit{et al.} 2003, Hennequin 2004,
di Stefano \textit{et al.} 2004), but an adequate choice of the experimental
parameters leads to a stable cloud, with atoms whose residual motion
is the thermal agitation.

A classical atom follows the motion equations of any classical object,
and in particular the Newton's second law $F=m\ddot{r}$, where $F$
is the force, $r$ the position and $m$ the mass (in the following,
we take $m=1$). When such an atom is dropped in a stationary wave,
it undergoes a force $F$, the potential $U$ of which is proportional
to the wave intensity $I$, and inversely proportional to the detuning
$\Delta$ between the wave frequency and the atomic transition frequency:\begin{eqnarray}
F & = & -\nabla U\\
U & \propto & \frac{I}{\Delta}\end{eqnarray}
Thus, atoms accumulate in bright (resp. dark) sites for $\Delta<0$
(resp. $\Delta>0$). When the atoms are cooled with the MOT, the atomic
density in these optical lattices is small enough to neglect the collisions
between atoms, and so the only source of dissipation is the spontaneous
emission. As spontaneous emission is proportional to $I/\Delta^{2}$,
it is relatively easy to build conservative optical lattices. Moreover,
the classical or quantum nature of the atoms in the lattice can be
adjusted continuously, as it depends on the ratio between the temperature
(or energy) of the atoms, and the depth of the lattice wells. For
wells deep enough as compared to the atom temperature, the quantum
properties of the atom, and in particular tunnelling, vanish, and
thus atoms can be considered as classical (Greiner 2001). In the following,
we always consider classical atoms, as discussed earlier.

The atom dynamics in the lattice depends on the dimensionality of
the lattice. For example, in a 1D lattice, atoms have only two dynamical
degrees of freedom, and thus even if the potential is not harmonic,
the dynamics cannot be complex. It is needed to add at least a periodic
forcing in such a lattice to observe chaos. On the contrary, a 2D
lattice can exhibit chaos, without external forcing.

\begin{figure}
\includegraphics[bb=0bp 0bp 811bp 415bp,clip,width=9cm]{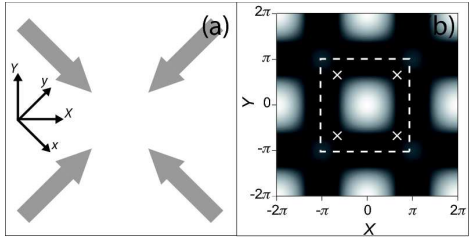}\label{Flo:fig1}

\caption{a) Layout of the laser beams. b) Spatial distribution of the intensity
in the $\left(X,Y\right)$ space. Black corresponds to the minimum
value (zero intensity), while white corresponds to the maximum. The
dotted square delimits the elementary mesh of the lattice, and the
white crosses are the saddle points. }

\end{figure}

But the atom dynamics also depends on the lattice geometry, and numerous
lattice geometries can be obtained, as e.g. a vertical stack of ring
traps (Courtade \textit{et al.} 2006), five-fold symmetric lattice
(Guidoni \& Verkerk 1999) or even quasiperiodic lattices (Guidoni
\textit{et al.} 1997). In this paper, we will focus on the case of
two orthogonal stationary plane waves with the same polarization.
The configuration of the laser beams is shown on Fig. 1a. The total
field is $\mathcal{E}=\cos kx+e^{i\phi}\sin ky$, where $x$ and $y$
are the two space coordinates, $\phi$ a phase, $k=2\pi/\lambda$
the wave vector and $\lambda$ the wavelength of the laser beam. The
intensity can be written as:\begin{equation}
I=\cos^{2}kx+\cos^{2}ky+2\alpha\cos kx\cos ky\end{equation}
where $\alpha=\cos\phi$. With the adequate normalization, the potential
is\begin{equation}
U_{\pm}=\pm I\end{equation}
where the explicit sign is that of $\Delta$. When $\alpha=0$, the
coupling between $x$ and $y$ disappears, and the problem becomes
separable. In all the other cases, the coupling between $x$ and $y$
could induce complex dynamics. It is easy to see that in these cases
the elementary mesh of the potential is turned of $\pi/4$ as compared
to the $\left(x,y\right)$ axes, and thus it is natural to introduce
the following new coordinates: \begin{eqnarray}
X & = & kx+ky\\
Y & = & ky-kx\end{eqnarray}
The intensity and the potential can now be written:\begin{equation}
I=U_{+}=-U_{-}=1+\alpha\left(\cos X+\cos Y\right)+\cos X\cos Y\end{equation}

Before studying the potential, let us concentrate on the intensity.
As an example, Fig. 1b shows the spatial distribution of the intensity
for $\alpha=0.5$. The elementary mesh is indicated through the dotted
line. Assuming $\alpha>0$, the intensity $I$ has an absolute maximum
$2\left(1+\alpha\right)$ at coordinates $\left(n2\pi,m2\pi\right)$,
where $m$ and $n$ are integers. It has also a relative maximum $2\left(1-\alpha\right)$
in $\left(\pi+n2\pi,\pi+m2\pi\right)$. Once again, we see that $\alpha=0$
is a special case because the absolute and relative maxima have the
same height. Note that $\alpha=1$ is another special case, where
the intensity at the relative maximum vanishes and, thus, is equal
to the minimum value. In this special case, we have black lines along
$X=\pi+n2\pi$ and $Y=\pi+n2\pi$. We will not consider these cases
in the following. On the other hand, the intensity goes to zero in
$\left(\pi+n2\pi,m2\pi\right)$ and $\left(n2\pi,\pi+m2\pi\right)$.
Two neighboring zeros are separated by a saddle point where the intensity
has the value $I=1-\alpha^{2}$. It is important to note that these
saddle points are on the bissectors, connecting on a straight line
an absolute maximum to a relative one and again to the next absolute
maximum. On the contrary, the saddle points do not stand on the straight
line that connect two neighboring zeros. This will induce a huge difference
in the dynamics of atoms in the lattice obtained for red detunings
($\Delta<0$), where the atoms are attracted in high intensity regions
and the one for blue detunings ($\Delta>0$), where the atoms are
repelled from these same regions. The bissectors are clearly escape
lines for the atoms when $\Delta<0$, while it is not the case for
$\Delta>0$.

Optical lattices appear to be an exciting tool to study the dynamics
of a conservative complex system, but how to characterize this dynamics
in the experiments? What are the experimentally accessible quantities?
The typical size of a lattice mesh is $\lambda/2$, i.e. 426 nm for
Cesium. As the diameter of a cold atom cloud is typically 2 mm, the
$10^{8}$ atoms are dropped in $22\,10^{6}$ sites for a 2D lattice,
which lead to 5 atoms/site. At these scales, it is clear that there
is no way to isolate an atom, and thus no way to follow its trajectory.
Moreover, to see an atom, we need light, and thus the measure introduces
a dissipation and destroy the atomic state. A typical measure consists
in illuminating the atoms with a laser flash, and recording the fluorescence
of the atoms through a camera. This destructive measure gives snapshots
of the atom distribution in the space. We examine in the following
if it is possible to extract informations about the atom dynamics
from this type of measurement.

\section{Dynamics of atoms inside the wells}

Before we search for signatures of the dynamics in the experimental
measurements, let us investigate in more details what are the relevant
parameters and characteristics of the atom dynamics in a lattice.
To illustrate this approach, let us consider again the two lattices
introduced in Section 2. Although these two lattices differ only by
the sign of their potential, they are deeply different. $U_{-}$ has
its wells where the light intensity is maximum, while $U_{+}$ has
its wells where the intensity vanishes. Let us denote $E_{T}$ the
value of the potential energy at the saddle point of the intensity.
Atoms, the energy $E$ of which is smaller than the threshold $E_{T}$,
are trapped into one site because they cannot climb up to the saddle
point. On the contrary, atoms with $E>E_{T}$ can travel between sites,
if they move in the good direction.

Inside a trap site, the energy of the atom plays the role of a stochastic
parameter. Indeed, for low energies, the atoms remain located close
to the bottom of the well, and their dynamics can be approximated
by an harmonic motion. As the energy increases, the potential becomes
more and more anharmonic, the nonlinearities increase, and the dynamics
can become more and more complex. To be able to compare the behavior
of atoms in different potentials, we take in the following the origin
of the energy at the bottom of the wells, and normalize the energy
so that $E_{T}=1$. The potential energy then takes a different form
for red and blue detunings:

\begin{equation}
U_{+}=\frac{I}{1-\alpha^{2}}\end{equation}

\begin{equation}
U_{-}=\frac{2\left(1+\alpha\right)-I}{\left(1+\alpha\right)^{2}}\end{equation}

Let us now examine in details the dynamics of the atoms in our two
potentials. The most relevant way is to look at the evolution of the
Poincaré sections as a function of the energy. Our phase space is
4-dimensional, with directions $\left(X,Y,\dot{X},\dot{Y}\right)$,
but because of the energy conservation, the accessible space reduces
to a 3D surface. We choose to consider Poincaré section at $\dot{Y}=0$
with increasing values, and thus, Poincaré sections are in the 3D
space $\left(X,Y,\dot{X}\right)$, and they lie on a 2D surface $S_{P}$,
which has the shape looking like a semi-ellipsoid. To represent the
Poincaré sections we can project them on the $\left(X,Y\right)$ plane
or on the more usual $\left(X,\dot{X}\right)$ plane. The latter shows
the Poincaré sections viewed from the vertex of the semi-ellipsoid.
However here, because of the stiff sides of $S_{P}$, the projection
in this plane leads to a confuse map, as many curves are projected
at the same location, and thus are superimposed. On the contrary,
the projection on the $\left(X,Y\right)$ plane gives more details,
and thus in the following, we often choose it. However, let's keep
in mind that we look at a lateral projection of a {}``bell'', and
thus that we superimpose its front and rear faces.

As pointed out before, because of the normalization we choose for
the energy, the form of the potential energy differs in the cases
of blue or red detuned lasers. We investigate each cases separately. 

In the case of red detuned lasers, the potential energy takes the
form :

\begin{eqnarray}
U_{-} & = & \omega_{0}^{2}\left(1-\cos X\right)+\omega_{0}^{2}\left(1-\cos Y\right)-\frac{\left(1-\cos X\right)\left(1-\cos Y\right)}{\left(1+\alpha\right)^{2}}\label{eq:potneg}\\
\mathrm{with\:}\omega_{0}^{2} & = & \left(1+\alpha\right)^{-1}\end{eqnarray}

This potential appears to be the sum of two simple pendula coupled
through the third term. The frequency for oscillations with small
amplitude is the same for the two directions. This degeneracy together
with the coupling term leads to a strong synchronization of the motion
in the two directions (Bennet \textit{et al.} 2002). However, in contrast
with the Huygens clocks, we do not have any dissipation here, so the
frequency locking occurs in a more subtle way (Hennequin \& Verkerk
2010). 

However, it is interesting to identify the resonances of the system.
A very simple approach is to restrict the problem to the first anharmonic
terms, similar to the undamped Duffing oscillator. We then look for
a periodic harmonic solution in the form $X=X_{0}\: cos\left(\omega t\right)$
and $Y=Y_{0}\: cos\left(\omega t+\varphi\right)$, with $\omega$
close to $\omega_{0}$. We drop terms at other frequencies (i.e. $3\omega$)
and we end up with six families of solutions. The first two are the
trivial ones : motion along the $X$ or the $Y$ directions ($Y_{0}=0$
or $X_{0}=0$). The other four are obtained for $X_{0}=Y_{0}$ and
for $\varphi=0,\:\pi,\:\pm\pi/2$. For a given energy $E$, the relations
giving $X_{0}$ and $\omega$ are not simple, and it is beyond the
aim of this article to write them explicitly. For the large amplitudes
considered in the following, the motion is not any more harmonic and
we cannot keep only the lower order terms, but the main result remains:
we have six periodic trajectories, leading to points in the Poincaré
section (except for the trajectory $Y_{0}=0$, that we cannot catch
in a Poincaré section at $\dot{Y}=0$). In the 3D space, these points
have the coordinates $\left(0,-Y_{0},0\right)$, $\left(\pm X_{0},-X_{0},0\right)$
and $\left(0,-X_{0}',\pm\omega X_{0}'\right)$.

\begin{figure}
\includegraphics[bb=0bp 0bp 493bp 501bp,width=12cm]{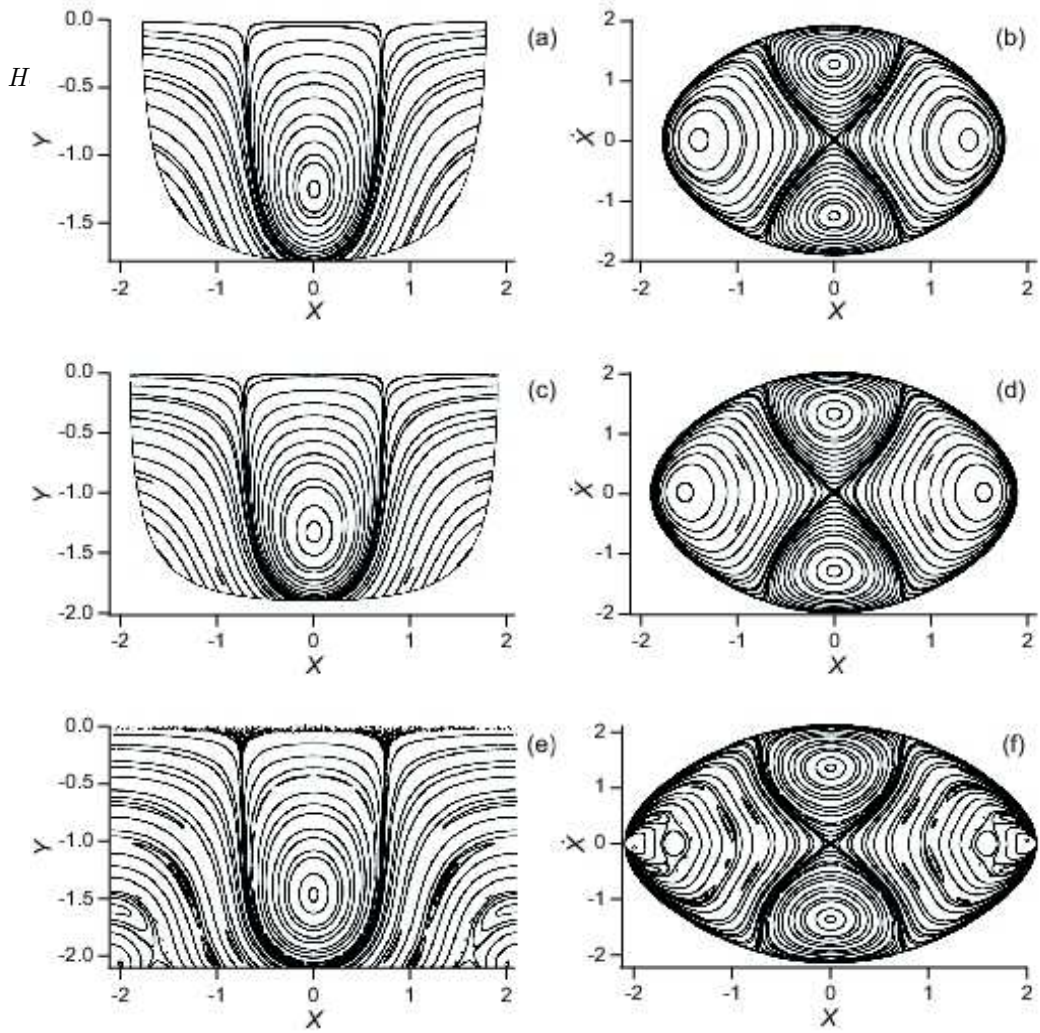}

\caption{$\left(X,Y\right)$ (left) and $\left(X,\dot{X}\right)$ (right) Poincaré
sections of the atomic dynamics in the $U_{-}$ potential. (a) and
(b): $E=0.80$; (c) and (d): $E=0.88$; (e) and (f): $E=1.00$.}

\end{figure}
In Fig. 2, we show the dynamics in the $U_{-}$ potential for different
normalized energies in the case of $\alpha=0.5$. These results have
been obtained through numerical resolution of the equations of motion
which are derived from the potential (\ref{eq:potneg}), without the
addition of any random quantity. All the described behaviors are thus
deterministic. For each value of the energy, we project the Poincaré
section on the $\left(X,Y\right)$ plane (left figures) and on the
$\left(X,\dot{X}\right)$ plane (right figures). For low enough energies
(e.g. $E=0.8$ in Fig. 2a and 2b), we see four distinct domains separated
by an X-shaped separatrix. In each of these domains, the Poincaré
section is cycling around one of the non-trivial resonances found
above. As the motion along $X$ and $Y$ is governed by the same frequency,
and because of the coupling between these two pendula, a synchronization
between the two directions occurs, through a phase locking between
the two motions. The corresponding behavior can be described as mainly
a $\omega$ periodic cycle perturbed by small sidebands (Hennequin
\& Verkerk 2010).

The dynamics in $U_{-}$ evolves only slightly when $E$ is increased.
The Poincaré surfaces are always organized around the separatrix delimiting
4 areas. In each area, the nature of the motion is the same, namely
phase locking between the motions in the $X$ and $Y$ directions.
Chaos appears close to the separatrix for $E\simeq0.88$ (Fig. 2c
and 2d), but it remains marginal, even when $E=1$ (Fig. 2e and 2f).
This very small extent is due to the original degeneracy of the frequencies
of the coupled pendula and to the strong coupling between them (Hennequin
\& Verkerk 2010).

\begin{figure}
\includegraphics[bb=0bp 0bp 529bp 489bp,width=12cm]{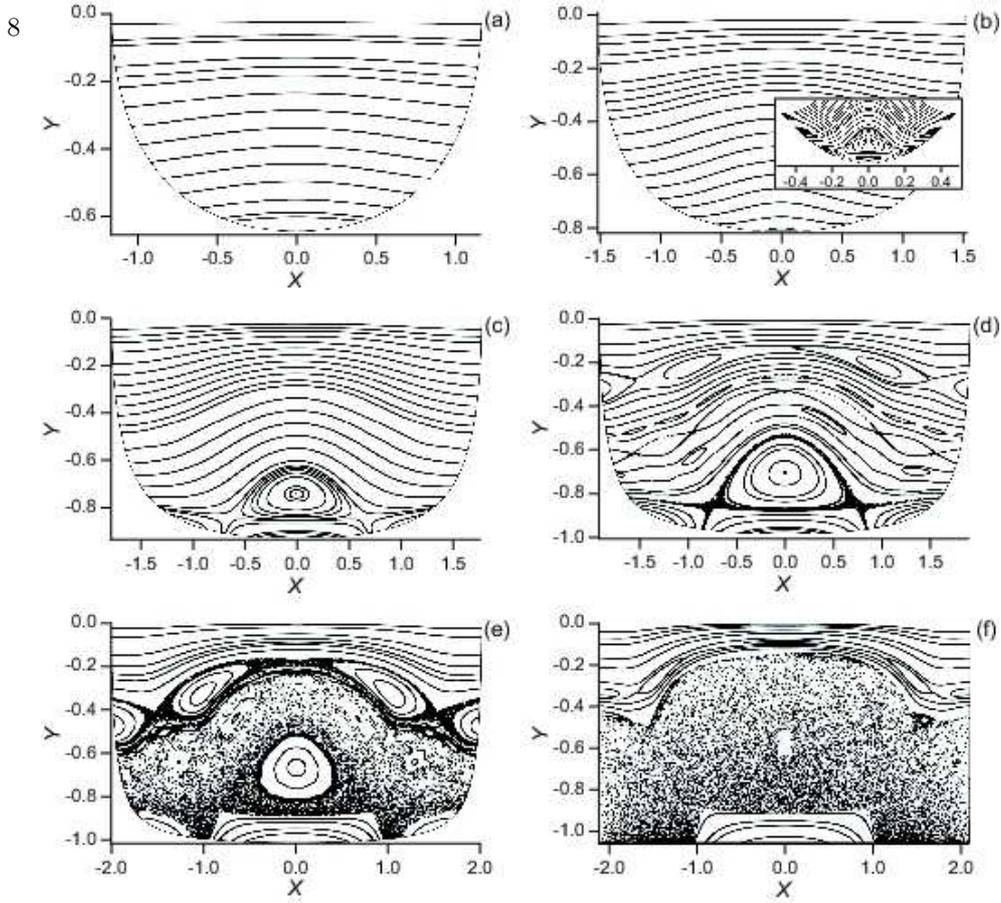}

\caption{$\left(X,Y\right)$ Poincaré sections of the atomic dynamics in the
$U_{+}$ potential. (a) $E=0.4$, (b) $E=0.63$, (c) $E=0.80$, (d)
$E=0.88$, (e) $E=0.93$ and (f) $E=1.00$.}

\end{figure}
For blue detunings ($\Delta>0$), the bottom of the well corresponds
to $I=0$, i.e. $\left(X=0,Y=\pi\right)$ sites. For the sake of simplicity,
we shift the origin in $Y$ by $\pi$, in order to have a trapped
motion centered at the origin. Thus, we can write :

\begin{eqnarray}
U_{+} & = & \omega_{0X}^{2}\left(1-\cos X\right)+\omega_{0Y}^{2}\left(1-\cos Y\right)-\frac{\left(1-\cos X\right)\left(1-\cos Y\right)}{\left(1-\alpha^{2}\right)}\\
\mathrm{with\;}\omega_{0X}^{2} & = & \left(1+\alpha\right)^{-1}\\
\omega_{0Y}^{2} & = & \left(1-\alpha\right)^{-1}\end{eqnarray}
Once again, this potential appears to be the sum of two coupled pendula.
But now, the two frequencies for oscillations with small amplitudes
are different: for the value $\alpha=0.5$ chosen here, the ratio
$\sqrt{3}$ of these two frequencies is irrational.

For very small energies (Fig. 3a), the dynamics consists essentially
in a regular motion around the bottom of the well, along a quasiperiodic
trajectory with frequencies $\omega_{X}$ and $\omega_{Y}$ close
to $\omega_{0X}$ and $\omega_{0Y}$. At the top of Fig. 3a, Poincaré
sections are those of atoms, the motion of which is essentially along
the $X$ axis. In $Y=0$, the trajectory is a periodic cycle along
the $X$ direction (edge of the semi-ellipsoid). At the opposite,
the periodic cycle at the bottom of the figure corresponds to the
situation where the atomic motion is exclusively along the $Y$ axis
(vertex of the semi-ellipsoid). Note that the nature of the motion
along these quasiperiodic cycles is deeply different from those described
with $\Delta<0$. Indeed, as $\omega_{0X}$ and $\omega_{0Y}$ are
very different, no locking occurs. In particular, in the spectrum
of the motion, the two main frequencies are close to $\omega_{0X}$
and $\omega_{0Y}$.

As the energy of the atom is increased, the atom can climb more and
more in the well, the frequencies $\omega_{X}$ and $\omega_{Y}$
change because of the anharmonicity of the potential, but the dynamics
does not change fundamentally until $E\simeq0.6$. At that point a
new feature appears: a stable periodic trajectory shows up as a cycle
close to the bottom of Fig. 3b, obtained for $E=0.63$. In fact, for
amplitudes large enough, the frequencies $\omega_{X}$ and $\omega_{Y}$
depart so much from their initial values $\omega_{0X}$ and $\omega_{0Y}$
that a new resonance appears at $\omega_{Y}=2\omega_{X}$.

For higher energies, the $\omega_{Y}=2\omega_{X}$ resonance grows
and comes closer to the centre of the figure and influence a non-negligible
fraction of the trajectories. In Fig. 3c, for $E=0.8$, the resonance
is clearly visible in $Y\simeq-0.74$. In the $\left(X,Y,\dot{X}\right)$
space, its Poincaré section consists in 2 points (superimposed in
the projection of fig. 3c), explored alternatively by the trajectory.
Around this point, the Poincaré sections are a double closed loop.
The corresponding quasiperiodic motion consists in a perturbed $\omega_{Y}=2\omega_{X}$
phase locked periodic cycle, where the perturbation consists in small
sidebands of $\omega_{X}$ and $\omega_{Y}$ in the spectrum. Thus
the separatrix appears here to be the limit between this phase locked
and the unlocked behaviors. The central domain and the two linked
lateral domains (bottom left and right) correspond to the phase locking.
The difference between these two domains is the relative phase on
the motion along $X$ and along $Y$. In the two other domains (top
and bottom), there is no locking between the $\omega_{X}$ and $\omega_{Y}$
frequencies.

In $E=0.8$ (Fig. 3c), all the trajectories are still periodic cycles
or quasiperiodic tori. When the energy is increased further, chaos
appears at $E\simeq0.88$, starting in the vicinity of the separatrix
(Fig. 3d). Then, it expands with some quasiperiodic islands remaining
(Fig. 3e), but finally, for $E=1$ (Fig. 3f), the only significant
quasiperiodic domains are those around the $X$ and $Y$ periodic
cycles. Around the locked periodic cycles, a narrow area with tori
remains, but chaos appears really to be dominant.

We have shown in this section that it is relatively easy to find two
slightly different lattices with fundamentally different dynamics.
These two configurations are easy to reach experimentally, as they
differ only by the sign of the detuning. It would be interesting now
to examine how to measure experimentally these differences, and if
these differences have an impact on the dynamics of atoms when they
jump between sites of the lattice. The next section deals with the
latter.

\section{Dynamics of atoms visiting several wells}

To travel from site to site, an atom needs to have an energy $E\geqslant1$,
but it is not a sufficient condition. Only atoms with an adequate
trajectory will effectively escape from a well. This implies that
for a given energy $E\geqslant1$, at least two classes of atoms can
exist: trapped atoms remaining in a single well, and traveling atoms,
which escape the wells. In fact, the situation is more complex, as
we will see now.

\begin{figure}
\includegraphics[scale=0.75]{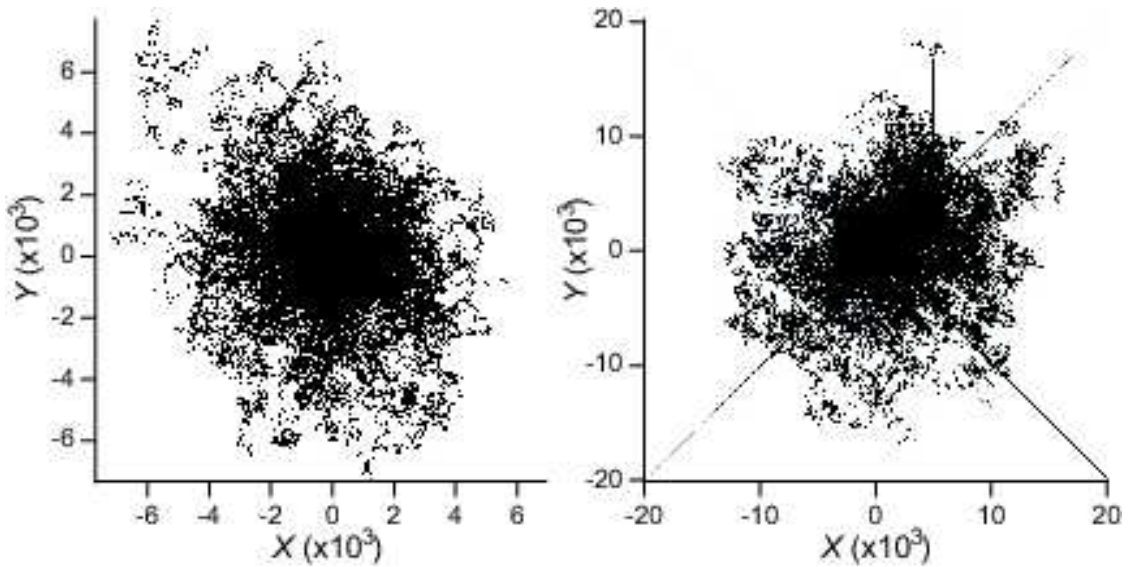}

\caption{$\left(X,Y\right)$ plot of the trajectories of 100 atoms in the (a)
$U_{+}$ ($E=2.66$) and (b) $U_{-}$ ($E=1.07$) lattices. Each atom
starts in the central mesh, and move during the time $t=10^{6}$,
which corresponds to more than $10^{5}$ periods of oscillation at
the bottom of a well.}

\end{figure}

Let us first examine the dynamics of traveling atoms in the blue case
($\Delta>0$). We are here interested by atoms with an energy $1<E<4$.
Indeed, atoms with $E>4$ have an energy larger than the potential
maximum, and thus they {}``fly'' above the potential, and their
trajectory is purely ballistic. On the contrary, the dynamics of the
atoms with an intermediate energy consists in complex trajectories
visiting a large number of sites, as in a random walk. As our model
is fully deterministic, it involves in fact chaotic trajectories.
Fig. 4a illustrates such a chaotic diffusion: it reports the trajectories
followed by 100 atoms. Such a trajectory is in fact an alternation
of oscillations inside wells and of jumps between wells. Here, we
know that chaos dominates inside the wells, and thus the chaotic nature
of the diffusion is not surprising. However, as we will see below,
the existence of chaos inside the wells is not a necessary condition
to observe a chaotic diffusion.

To think of an experimental characterization of this chaotic diffusion,
a simple way would be to characterize the diffusion function, and
to evaluate a diffusion coefficient. Fig. 5a reports the distance
covered by atoms of high energy ($E=2.66$) as a function of time.
With such high energies, all atoms travel between wells. They all
follow a similar behavior, characterized by a diffusion over a distance
of the order of $10^{3}$ for the time interval of the figure. Although
there is a small dependence of these curves as a function of the energy
of the atom, the orders of magnitude remain the same for all energies
$1<E<4$. The only difference is that for lower energies, some atoms
remain trapped in their well, and so a second group of curves appears
with atoms remaining within a short distance (smaller than the mesh,
i.e. $2\pi$) of their initial location.

In the red detuned situation ($\Delta<0$), the maximum of the potential
is at $E=1.33$. As in the blue case, atoms with an energy $E>1.33$
have ballistic trajectories, and atoms with $1<E<1.33$ exhibit a
diffusive chaotic behavior (Fig. 4b). The origin of chaos now is clearly
in the jumps between wells, as the dynamics in the wells is regular.
And in fact, there is a main difference as compared to the blue case:
the diffusion scale is larger by one order of magnitude, on the whole
interval $1<E<1.33$. We did not check if the slower diffusion originates
effectively in the chaotic trajectories followed by the atoms inside
the wells, but it would be interesting to check in a future work how
these chaotic behavior could slow down the atoms. However, the difference
of one order of magnitude in the diffusion speed reveals that the
macroscopic behavior of atoms could effectively be used to characterize
the nature of the dynamics in optical lattices

But there is another important difference between the two lattices:
in the red case a third regime exists, neither trapping neither diffusing.
It is illustrated on Fig. 4b, where trajectories appear following
the two bissectors. These trajectories correspond to atoms traveling
along the escape lines of the lattice, as they were described in section
2. These atoms follow in fact a ballistic trajectory, where they travel
very rapidly along the bissectors. For example, in Fig. 4b, the ballistic
trajectories reach $10^{6}$ in all directions, while the diffusive
atoms reach only $2\,10^{4}$ of the same units in the same time.
Note that the ballistic trajectories we discuss here occur as soon
as the threshold $E=1$ is reach, and only along the escape lines
of the potential.

Fig. 5b shows the distance covered by the atoms as a function of time.
We have now clearly three groups of trajectories: trapped trajectories
at bottom, diffusive trajectories for distances of about $10^{4}$,
and ballistic trajectories at the top, for distances larger than $10^{5}$.
The main difference as compared to the $\Delta>0$ case is the cohabitation
of ballistic and diffusive trajectories, even just above threshold.
This put in evidence three specific time scales of the dynamics of
atoms with a given energy, associated respectively with the trapped,
the chaotic diffusive and the ballistic trajectories.

\begin{figure}
\includegraphics[width=11cm]{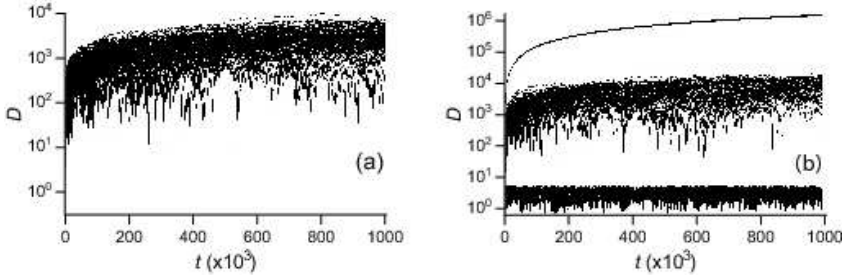}

\caption{Distance covered by 100 atoms as a function of time. Parameters are
the same as in Fig. 4: in (a), $U_{+}$ and $E=2.66$; in (b), $U_{-}$
and $E=1.07$.}

\end{figure}

In this section, we examined the dynamics of atoms, whose energy is
large enough to escape the potential wells, but remains smaller than
the potential maxima. We focused on atoms traveling between wells,
and found a different behavior for our two lattices. For the red lattice,
atoms can be classified following two types of dynamics: the diffusive
atoms exhibit a chaotic dynamics carrying them off their initial location;
the ballistic atoms move away rapidly from their initial location.
These behaviors are associated with two different time scale. But
is it sufficient to identify these different regimes in a real experiment?
We have also shown that the dynamics of atoms in the blue lattice
is quite different, both for the diffusive regime and the ballistic
one: the time scale of the former is one order of magnitude smaller,
while the latter simply does not exist. Can we use these properties
to characterize and distinguish experimentally the two lattices? These
questions are discussed in the next section.

\section{Macroscopic signatures of chaos}

Our aim is to characterize the dynamics of the cold atoms in the optical
lattice. As we are concerned by conservative lattices, we cannot hope
to {}``film'' in real time the atoms in the lattice, as it would
introduce dissipation. Thus we have to find other techniques. As the
specificity of each lattice concerns the traveling atoms, an experimental
measurement aiming at characterizing these lattices should characterize
these traveling atoms.

Experimentally, the lattice is finite. So the traveling atoms will
reach the edge of the lattice, and finally leave the lattice. Therefore
a simple measure of the lifetime of the atoms in the lattice give
informations about the trapped and traveling atoms. However, as there
are several types of traveling atoms, the simple measure of a lifetime
is not sufficient, and the lifetime curve itself, in particular its
shape, must be analyzed. Thus we will plot now the number of atoms
in the lattice as a function of the time. The shape of the curve and
the lifetime itself should give informations about the traveling atoms,
while the baseline gives the percentage of trapped atoms.

In the experiment, all the atoms have not the same energy, but on
the contrary, they exhibit a distribution of energies linked to their
temperature. Thus the results shown below have been obtained by using
a sample of atoms with an appropriate distribution of energy.

\begin{figure}
\includegraphics[width=11cm]{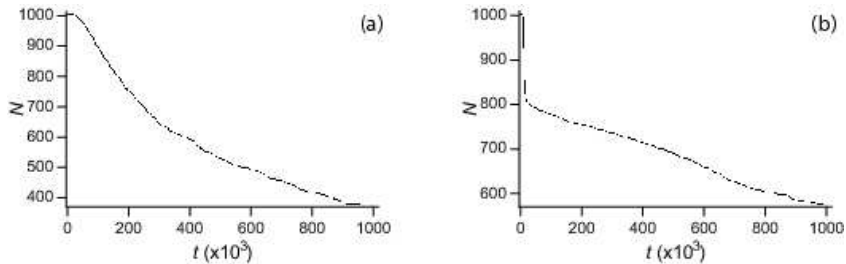}

\caption{Number of atoms versus time in (a) $U_{+}$ potential and (b) $U_{-}$
potential.}

\end{figure}
Fig. 6a shows the number of atoms in the blue lattice versus time.
To simulate the finite size of the lattice, atoms are removed as soon
as they reach a distance $D_{L}=1000$. The curve exhibits a plateau
at short times, followed by a exponential-like decreasing to an asymptote.
The plateau corresponds to the time needed by the first atoms to reach
the edge of the lattice (in the simulations, all the atoms are supposed
to be initially at the center of the lattice). The decreasing corresponds
to the diffusing atoms escaping the lattice, and the asymptote to
the number of atoms trapped in wells. This behavior does not depend
on the lattice size $D_{L}$, except that the life time of atoms increases.
In fact, the distance $D_{L}=1000$, i.e. about 150 lattice meshes
or 70 \textmu{}m for a Cs trap, is smaller by one order of magnitude
than a typical experimental realization. However, a value of $D_{L}=10^{4}$
leads, for the data presented in Fig. 4a and 5a, to an almost flat
curve, because the time series are not long enough. To reach such
a distance, one should increase the evolution time by two orders of
magnitude.

Fig. 6b shows the number of atoms in the red lattice versus time,
for $D_{L}=10^{4}$. The shape of the curve is qualitatively different
from that obtained for the blue lattice. At short times, a fast decreasing
appears, corresponding to the loss of the ballistic atoms. At large
times, not visible on the figure, an asymptote is reached, corresponding
to the trapped atoms. The intermediate decreasing correspond to the
loss of the diffusing atoms. Note that the decreasing appears to be
more or less linear. In fact, the shape of this part of the curves
is the sum of the diffusing losses of different classes of atoms differing
by their energy. As a function of $D_{L}$, this sum can exhibit very
different shapes, from an exponential-like shape, as in Fig. 6a, to
an almost linear shape, as in Fig. 6b.

Fig. 6 shows that the measure of the lifetime of atoms in a conservative
optical lattice provides qualitative and quantitative informations
about the nature of the lattice and the nature of the dynamics of
the atoms in the lattice, in particular about the chaotic diffusion.
Therefore the measure of the atom lifetime, in particular the existence
of several characteristic times in the decreasing of the atom number,
appears to be a signature of the chaotic dynamics of atoms in the
lattice.

\section{Conclusion}

We have shown in this paper that optical lattices are a good toy model
to study experimentally the dynamics of conservative systems, provided
that relevant experimental measures are found to characterize this
dynamics. In particular, we show that changing a simple experimental
parameter can lead to two deeply different lattices, where atoms exhibit
very different dynamical behaviors. We have shown that these differences
exist both in the local dynamics of atoms inside a well, and in the
non local dynamics of atoms traveling between wells. We searched numerically
for signatures of these different dynamics in the experimentally accessible
quantities, and found that the measure of atom lifetimes in the lattice
gives numerous informations about the existence and the type of chaotic
diffusion of the atoms.

It would be interesting now to characterize more precisely the diffusion
function, as a function of the experimental parameters, in particular
the atom temperature and the lattice size, and obviously to test these
results on a real experiment.

We considered in this paper only the dynamics of classical atoms,
and it appears that this dynamics is more complex and more subtle
than usually considered. Simple statistical analyses are not enough
to fully characterize this dynamics, and more suitable tools are necessary.
It is important now to think about the consequences of these results
in the quantum regime, and in particular, to what could be the equivalent
measures in the quantum world.

We thank Frédéric Carlier for fruitful discussions. The Laboratoire
de Physique des Lasers, Atomes et Molécules is Unité Mixte de Recherche
de l'Université de Lille 1 et du CNRS (UMR 8523). The Centre d'\'Etudes
et de Recherches Lasers et Applications (CERLA) is supported by the
Ministère chargé de la Recherche, the Région Nord-Pas de Calais and
the Fonds Européen de Développement \'Economique des Régions.

\section*{References}

\noindent {\footnotesize Bennett M., Schatz M. F., Rockwood H., and
Wiesenfeld K., Huygens\textquoteright{} clocks, Proc. Roy. Soc. London
A }\textbf{\footnotesize 458}{\footnotesize , 563-579 (2002).}{\footnotesize \par}

\noindent {\footnotesize Billy J., Josse V., Zuo Z. C., Bernard A.,
Hambrecht B., Lugan P., Clement D., Sanchez-Palencia L., Bouyer P.
and Aspect A., Direct observation of Anderson localization of matter
waves in a controlled disorder, Nature }\textbf{\footnotesize 453}{\footnotesize{}
891-894 (2008)}{\footnotesize \par}

\noindent {\footnotesize Chabe J., Lemarie G., Gremaud B., Delande
D., Szriftgiser P., and Garreau J. C., Experimental observation of
the Anderson metal insulator transition with atomic matter waves,
Phys. Rev. Lett. (2009)}{\footnotesize \par}

\noindent {\footnotesize Courtade E., Houde O., Clément J.-F., Verkerk
P. and Hennequin D., Realization of an optical lattice of ring traps,
Phys. Rev. A, }\textbf{\footnotesize 74}{\footnotesize{} 031403(R)
(2006)}{\footnotesize \par}

\noindent {\footnotesize di Stefano A., Fauquembergue M., Verkerk
P. and Hennequin D., Giant oscillations in the magneto-optical trap,
Phys. Rev. A }\textbf{\footnotesize 67}{\footnotesize{} 033404 (2003)}{\footnotesize \par}

\noindent {\footnotesize di Stefano A., Verkerk P. and Hennequin D.,
Deterministic instabilities in the magneto-optical trap, Eur. Phys.
J. D }\textbf{\footnotesize 30}{\footnotesize{} 243-258 (2004)}{\footnotesize \par}

\noindent {\footnotesize Douglas P., Bergamini S. and Renzoni F.,
Tunable Tsallis Distributions in Dissipative Optical Lattices, Phys.
Rev. Lett. }\textbf{\footnotesize 96}{\footnotesize , 110601 (2006)}{\footnotesize \par}

\noindent {\footnotesize Fang J. and Hai W., Wannier\textendash{}Stark
chaos of a Bose\textendash{}Einstein condensate in 1D optical lattices,
Physica B 370, 61 (2005)}{\footnotesize \par}

\noindent {\footnotesize Greiner M., Bloch I., Mandel O., Hänsch T.
W., and Esslinger T., Exploring Phase Coherence in a 2D Lattice of
Bose-Einstein Condensates, Phys. Rev. Lett. }\textbf{\footnotesize 87}{\footnotesize ,
160405 (2001)}{\footnotesize \par}

\noindent {\footnotesize Greiner M., Mandel O., Esslinger T., Hänsch
T. W. and Bloch I.,Quantum phase transition from a superfluid to a
Mott insulator in a gas of ultracold atoms, Nature }\textbf{\footnotesize 415}{\footnotesize ,
39 (2002)}{\footnotesize \par}

\noindent {\footnotesize Guidoni L., Triche C., Verkerk P. and Grynberg
G., Quasiperiodic optical lattices, Phys. Rev. Lett. }\textbf{\footnotesize 79}{\footnotesize ,
3363 (1997)}{\footnotesize \par}

\noindent {\footnotesize Guidoni L. and Verkerk Ph., Optical lattices:
cold atoms ordered by light, J. Opt. B: Quantum Semiclass. Opt. }\textbf{\footnotesize 1}{\footnotesize ,
R23 (1999)}{\footnotesize \par}

\noindent {\footnotesize Guidoni L., Depret B., di Stefano A., and
Verkerk P., Atomic diffusion in an optical quasicrystal with five-fold
symmetry, Phys. Rev. A }\textbf{\footnotesize 60}{\footnotesize ,
R4233 (1999)}{\footnotesize \par}

\noindent {\footnotesize Hennequin D., Stochastic dynamics of the
magneto-optical trap, Eur. Phys. J. D }\textbf{\footnotesize 28}{\footnotesize{}
135-147 (2004)}{\footnotesize \par}

\noindent {\footnotesize Hennequin D. and Verkerk P., Synchronization
in non dissipative optical lattices, Eur. Phys. J. D, DOI: 10.1140/epjd/e2009-00324-1
(2010)}{\footnotesize \par}

\noindent {\footnotesize Jaksch D. and Zoller P.,The cold atom Hubbard
toolbox, Ann. Phys. }\textbf{\footnotesize 315}{\footnotesize , 52
(2005)}{\footnotesize \par}

\noindent {\footnotesize Jersblad J. , Ellmann H., Stchkel K., Kastberg
A., Sanchez-Palencia L., and Kaiser R., Non-Gaussian velocity distributions
in optical lattices, Phys. Rev. A }\textbf{\footnotesize 69}{\footnotesize{}
013410 (2004)}{\footnotesize \par}

\noindent {\footnotesize Kuan W. H., Jiang T. F., and Cheng S. C.,
Instabilities of Bose-Einstein Condensates in Optical Lattices, Chin.
J. Phys. 45 219 (2007)}{\footnotesize \par}

\noindent {\footnotesize Lignier H., Chabé J., Delande D., Garreau
J. C. and Szriftgiser P., Reversible destruction of dynamical localization,
Phys. Rev. Lett }\textbf{\footnotesize 95}{\footnotesize , 234101
(2005)}{\footnotesize \par}

\noindent {\footnotesize Mandel O., Greiner M., Widera A., Rom T.,
Hänsch T. W. and Bloch I., Controlled collisions for multi-particle
entanglement of optically trapped atoms, Nature }\textbf{\footnotesize 425}{\footnotesize ,
937 (2003)}{\footnotesize \par}

\noindent {\footnotesize Paredes B., Widera A., Murg V., Mandel O.,
Fölling S., Cirac I., Shlyapnikov G. V., Hänsch T. W. and Bloch I.,
Tonks-Girardeau gas of ultracold atoms in an optical lattice, Nature
}\textbf{\footnotesize 429}{\footnotesize , 277 (2004)}{\footnotesize \par}

\noindent {\footnotesize Roati G., D'Errico C., Fallani L., Fattori
M., Fort C., Zaccanti M., Modugno G., Modugno M. and Inguscio M.,
Anderson localization of a non-interacting Bose\textendash{}Einstein
condensate, Nature }\textbf{\footnotesize 453}{\footnotesize , 895
(2008)}{\footnotesize \par}

\noindent {\footnotesize Steck D. A., Milner V., Oskay W. H., and
Raizen M. G., Quantitative Study of Amplitude Noise Effects on Dynamical
Localization, Phys. Rev. E }\textbf{\footnotesize 62}{\footnotesize ,
3461 (2000)}{\footnotesize \par}

\noindent {\footnotesize Thommen Q., Garreau J. C., and Zehnlé V.,
Classical Chaos with Bose-Einstein Condensates in Tilted Optical Lattices,
Phys. Rev. Lett. 91, 210405 (2003)}{\footnotesize \par}

\noindent {\footnotesize Vollbrecht K. G. H., Solano E. and Cirac
J. I., Ensemble quantum computation with atoms in periodic potentials,
Phys. Rev. Lett. }\textbf{\footnotesize 93}{\footnotesize , 220502
(2004)}{\footnotesize \par}

\noindent {\footnotesize Wilkowski D., Ringot J., Hennequin D. and
Garreau J. C., Instabilities in a magneto-optical trap: noise-induced
dynamics in an atomic system, Phys. Rev. Lett. }\textbf{\footnotesize 85}{\footnotesize{}
1839-1842 (2000)}{\footnotesize \par}

\end{document}